\documentclass[conference]{IEEEtran}
\IEEEoverridecommandlockouts
\usepackage{cite}
\usepackage{amsmath,amssymb,amsfonts,bm}
\usepackage{graphicx}
\usepackage{subfigure}
\usepackage{textcomp}

\usepackage{algorithm}
\usepackage{algpseudocode}
\usepackage{amsmath}
\def\BibTeX{{\rm B\kern-.05em{\sc i\kern-.025em b}\kern-.08em
    T\kern-.1667em\lower.7ex\hbox{E}\kern-.125emX}}

\begin{document}

\title{Suspension Regulation  of Medium-low-speed Maglev Trains via Deep Reinforcement Learning\thanks{F. Zhao, K. You and S. Song are with the Department of Automation and BNRist, Tsinghua University, Beijing 100084, China. e-mail: zhaofr18@mails.tsinghua.edu.cn, \{youky, shijis\}@tsinghua.edu.cn.}\thanks{W. Zhang and L. Tong are with the CRRC Zhuzhou Locomotive Co.  Ltd. e-mail: \{zhang-wen-yue,alanatlsh\}@126.com.}} 
%\thanks{Identify applicable funding agency here. If none, delete this.}}

\author{\IEEEauthorblockN{Feiran Zhao, Keyou You, Shiji Song, Wenyue Zhang, Laisheng Tong}}

\maketitle

\begin{abstract} The suspension regulation is critical to the operation of medium-low-speed maglev trains (mlsMTs).  Due to uncertain environment, strong disturbances and high nonlinearity of the system dynamics, this problem cannot be well solved by most of the model-based controllers. In this paper, we propose a sample-based controller by reformulating it as a continuous-state, continuous-action Markov decision process (MDP) with unknown transition probabilities. With the deterministic policy gradient and neural network approximation, we design reinforcement learning (RL) algorithms to solve the MDP and obtain a state-feedback controller  by using sampled data from the magnetic levitation system. To further improve its performance, we adopt a double Q-learning scheme for learning the regulation controller. We illustrate that the proposed controllers outperform the existing PID controller with a real dataset from the mlsMT in Changsha, China and is even comparable to model-based controllers, which assume that the complete information of the model is known, via simulations.

\end{abstract}


\section{Introduction}\label{sec_intro}
\begin{figure}[t]
	\centerline{\includegraphics[width=70mm]{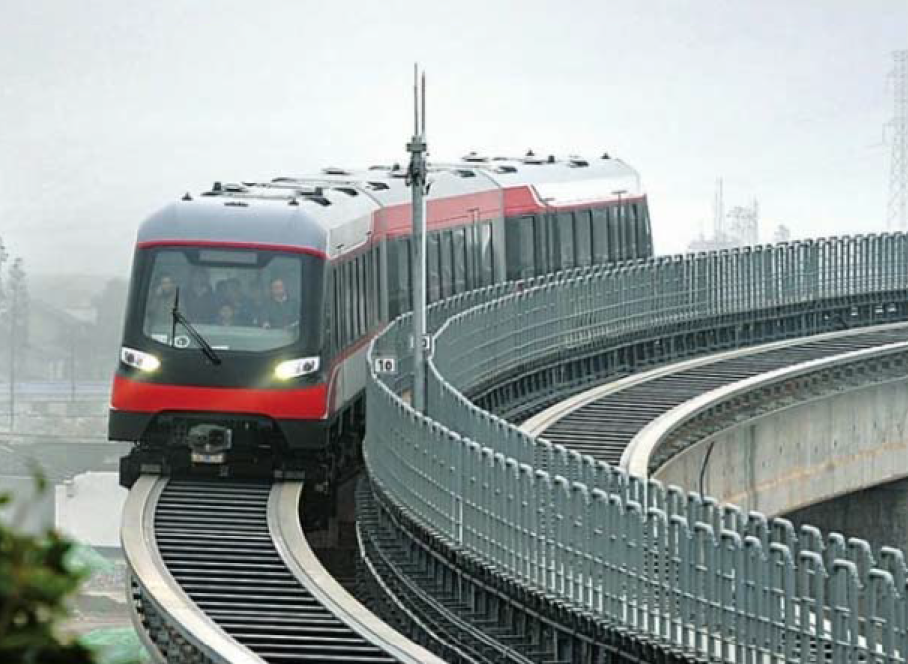}}
	\caption{The medium-low-speed maglev train in Changsha.}
	\label{picture}
\end{figure}

As one of the most competitive green ground transportation systems, the medium-low-speed maglev train (mlsMT), as shown in Fig. \ref{picture}, has advantages of low noise, small turning radius and high climbing ability, which are important to the city and intercity public transportation~\cite{lee2006review, boldea1999linear}. In China, the first medium-low-speed maglev rail line was put into commercial operation in 2016, connecting the CBD of Changsha to the Huanghua international airport. The regulation of the magnetic levitation system is one of the most fundamental problems for the operation of mlsMTs \cite{chen2016parallel, wu2008neural}. For example, a good controller can ensure the air gap  between the coil and the track stable even under large disturbances and provide passengers a smooth ride without severe bumps. In the Changsha's mlsMT rail line, it is desired to regulate this air gap to 8mm.
The main difficulties for regulating the magnetic levitation system of mlsMTs at least include: (a) An exact dynamical model of the magnetic levitation system is impossible to obtain as it usually contains unknown time-varying parameters, which depend not only on the working load but also the  environment. Hence, many advanced model-based controllers cannot be effectively applied. (b) The dynamical model  is overly complicated as it involves the coupling between the electromagnetic and kinematic equations, which renders that most of the linearization based controllers, e.g., the proportional-integral-derivative (PID) controllers, may not perform well.  (c) The operating environment, e.g., weather conditions, railway tracks and etc,  impose strong disturbance to the system.  

Most of the conventional control methods are mainly  based on an exact dynamical model. In~\cite{yadav2016optimized, kumar2013lqr}, an optimized PID controller is designed based on a linearized model at the equilibrium point. Although it outperforms the classical tuned PID controller, the performance may significantly degrade with an increasing deviation from nominal operating points.  In~\cite{torres2012feedback, vsuster2012modeling, morales2011nonlinear}, a nonlinear controller is designed by using feedback linearization. Robust control~\cite{yang2010robust, bonivento2005balanced} and low-bias control~\cite{tsiotras2005low} are also utilized for the regulation problem. However, they require an exact physical model of the system which is not practical. In~\cite{lin2011sopc}, a fuzzy-PID controller is designed by using the fuzzy inference for self-regulating PID parameters, yet its implementation largely depends on empirical knowledge. The PID controller can be also implemented without model information, while it requires empirical efforts for tuning parameters in the present problem, and even with optimal parameters, its performance remains unsatisfying.

Other model-based methods are proposed by using  a simplified model with estimated parameters. In~\cite{de2017modeling}, the controller is devised on an approximated model via Taylors series expansion around one operation point and then neural networks are utilized to approximate the unknown parameters. In~\cite{pradhan2016nonlinear}, an input-output feedback linearization method is proposed with the estimate of unknown parameters estimated by using the curve fitting of the real-time data, while the influence of external disturbances is neglected in the analysis. In~\cite{qin2014modeling}, a model predictive controller is designed with identified model parameters by applying the structured nonlinear parameter optimization method. Since the unknown parameters are identified offline, the performance may degrade due to the large variation of parameters.

Overall, the performance of the above mentioned methods  depends on the model accuracy and may significantly degrade due to the unmodeled dynamics, the external disturbances and the variation of parameters. Thus a sample-based controller is essential. Notice that the mlsMT commutes in the same rail, we solve the regulation problem by designing a reinforcement learning (RL) based controller in this paper.  Under the dynamic programming framework, the RL aims to solve Markov decision process (MDP) problems by solely using training samples of the decision process.  It is a model-free method, since the controller is not built directly on the knowledge of the dynamical model. It has been applied successfully in the control domain, including the traffic signal control \cite{chu2019multi, el2013multiagent}, the train operation control \cite{yin2014intelligent}, the path planning for a mobile robot~\cite{konar2013deterministic} or multirobot~\cite{rakshit2013realization}, the multirobot cooperation~\cite{la2014multirobot},  the autonomous underwater vehicle (AUV)~\cite{wu2018depth}, etc. 

Usually, the performance of the RL heavily  depends on the quality and quantity of training samples. This is not a problem in the regulation problem of this work, whose objective is to control the air gap between the coil and the track to the desired level, as the mlsMT always runs in the same rail and the training data is easily recyclable. Since an accurate dynamical model of the magnetic levitation system is difficult to obtain in practice, a sample-based RL algorithm is designed to learn a controller from samples of the magnetic levitation system of mlsMTs.

The key of designing RL based controller is how to properly reformulate the regulation problem as an effective MDP, where an agent takes an action, which is the control input to the magnetic levitation system, and the current state transits to the next state with an one-step cost. In the regulation problem, a good design of state and one-step cost for the MDP can immensely improve the performance of RL algorithms. A typical choice of the state is the state variables of the magnetic levitation system, including the regulation error and its derivative. The one-step cost is naturally defined as the combination of the regulation error and the energy cost. This selection of the state and the one-step cost has been verified in this paper.

Thus, the regulation of the magnetic levitation system is converted to solve a well-defined continuous-state, continuous-action MDP with unknown transition probabilities. Moreover, we design actor-critic neural networks to parameterize the value function and the policy function of the MDP, respectively. Then, we use  the deterministic policy gradient~\cite{silver2014deterministic} with experience replay to train networks, which forms the Deep Deterministic Policy Gradient algorithm (DDPG) of this work, and obtain a state-feedback controller for the suspension regulation.

To further improve the control performance, we adopt a double Q-learning scheme \cite{fujimoto2018addressing} and derive a Twin Delayed Deep Deterministic Policy Gradient algorithm (TD3). We further discuss the implementation of the RL algorithms on the regulation problem, and propose an initialization method to accelerate the convergence by learning a real dataset from the existing controller. We illustrate with simulations that our RL based controller is comparable to model-based control methods, which rely on the complete information of the magnetic levitation system. We also test our controllers on a real dataset sampled from the Changsha's mlsMT rail line, and the results show that our RL framework performs well even with a limited amount of data. More importantly, the proposed controller improves the existing PID controller in the Changsha's mlsMT. 

The main contributions of this paper are summarized as follows.
\begin{enumerate}
\renewcommand{\labelenumi}{\rm(\alph{enumi})} 
\item We are the first to solve the regulation problem of the magnetic levitation system with the deep reinforcement learning based controller.
%\item Two structures of neural networks are designed as the actor network and the critic network.
\item  We adopt a double Q-learning scheme to address the function approximation error in actor-critic networks.
\item  We propose an initialization method to learn the real dataset to reduce the sample complexity and improve the performance.
\end{enumerate}

The reminder of this paper is recognized as follows. In Section \ref{sec_problem}, we formulate the control problem with an approximate nonlinear dynamical model. In Section \ref{sec_mdp}, we model the suspension regulation problem as a MDP. In Section \ref{sec_ddpg}, the RL framework is adopted to solve the MDP and the structures of two neural networks are designed. In Section \ref{sec_td3}, we adopt a double Q-learning scheme to reduce the approximation error of neural networks. In Section \ref{sec_alg}, we discuss implementation of the RL framework on the suspension regulation problem and provide an initialization method to accelerate the convergence. In Section \ref{sec_exp},  simulations are conducted and the performance of the model-based methods are compared to show the effectiveness of RL algorithms. In addition, we perform an experiment on a real dataset to demonstrate the advantage  of the RL controller.

\section{Problem Formulation}\label{sec_problem}
In this section, we describe the magnetic levitation system of the mlsMT in details.

\subsection{The Regulation Problem}
\begin{figure}[t]
\centerline{\includegraphics[width=70mm]{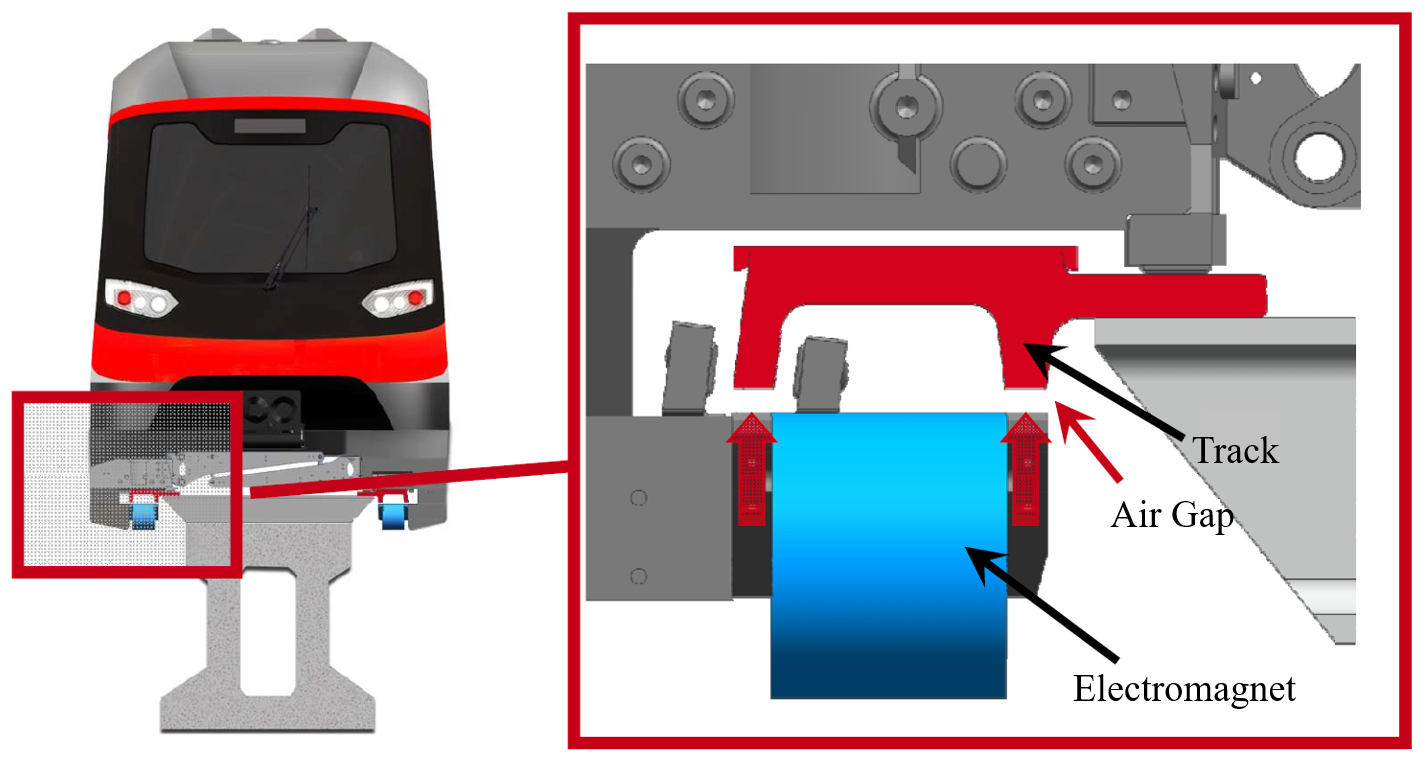}}
\caption{An overview of the magnetic levitation system.}
\label{lsm}
\end{figure}

\begin{figure}[t]
\centerline{\includegraphics[width=70mm]{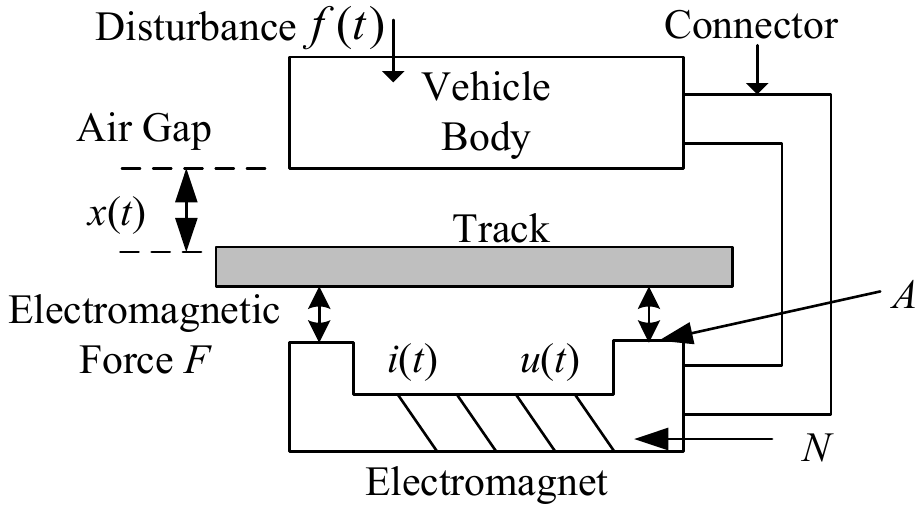}}
\caption{The illustrative magnetic levitation system.}
\label{sus}
\end{figure}

The suspension of the mlsMT is achieved by the attraction force between the rail track and the electromagnets embedded on the bottom of the vehicle body, see Fig. \ref{lsm} for an illustration. In operation, it is essential to regulate the air gap between the rail track and the electromagnets to a desired level, which is $8$mm in the Changsha's mlsMT rail line. To solve such a regulation problem,  the voltage in the coil of each electromagnet needs to be carefully controlled.  Usually, the mlsMT consists of a series of electromagnets to cooperatively suspend the train and  the voltage control of all the coils is coupled, which is too complicated for design.  In this work,  their coupling is neglected and the regulation problem is simplified as the voltage control of the coil of a single electromagnet, which is shown in Fig. \ref{sus}.

In Fig. \ref{sus}, $F$ represents the electromagnetic force between the electromagnet and the track. The alphabet $A$ denotes the effective magnetic pole area of the electromagnet. $N$ denotes the number of turns of the coil. $i(t)$ and $u(t)$ denote the current and voltage in the coil of an electromagnet, respectively. $f(t)$ represents the disturbance, which might be mainly contributed by the mass of passengers. $R$ denotes the coil resistance.  For the model-based approach, it is required to exactly describe the relationship between the above quantities, which is clearly impossible in practice, e.g., the effective magnetic pole area $A$ is usually not accessible and depends on the relative position between the mlsMT and the electromagnet. In the literature, an approximate nonlinear model has been widely adopted and is given in the next subsection.

\subsection{An Approximate Nonlinear Dynamical Model}
Before preceding, it is necessary to provide an intuitive understanding on the dynamics of the magnetic levitation system. Note that an accurate dynamical model is difficult to obtain in practice, thus we only present an approximate dynamical model~\cite{sun2016design}.

Assume that the mlsMT is in the equilibrium state, which includes the desired air gap $x_{eq}$, the current in the coil $i_{eq}$ and the input voltage $u_{eq}$.  Let \begin{equation}
z(t)=
\left[
\begin{array}{ccc}
    z_1(t) \\
    z_2(t) \\
    z_3(t)
\end{array}
\right]
=
\left[
\begin{array}{ccc}
    x(t)-x_{eq} \\
    \dot{x}(t) \\
    i(t) - i_{eq}
\end{array}
\right],
\label{state_sys}
\end{equation} 
where $\dot{x}(t)$ denotes the derivative of $x(t)$ with respect to time. Then, the following nonlinear model for $z(t)$  has been widely adopted \cite{sun2016design}, i.e., 
\begin{equation}
\begin{aligned}
\dot{z}_1(t)&=  z_2(t), \\
\dot{z}_2(t) &=  \frac{(2x_{eq}+z_1(t))z_1(t)g}{(z_1(t)+x_{eq})^2}\\
&~~~~~ -\frac{(2\sqrt{\kappa}x_{eq}+z_3(t))z_3(t)g}{\kappa(z_1(t)+x_{eq})^2}+\frac{f(t)}{m},   \\
\dot{z}_3(t)&=   \frac{z_3(t)+\sqrt{\kappa}x_{eq}}{z_1(t)+x_{eq}}z_2(t)
+ \frac{\kappa(z_1(t)+x_{eq})}{2mg}\\ & ~~~~~ \times (\upsilon(t)-z_3(t)R),
\end{aligned}
\label{dynamic}
\end{equation}
where $\kappa = {4mg}/{(\mu_0N^2A)}$ is assumed to be a constant, $m$ denotes the mass of the train, and $\upsilon(t)=u(t)-u_{eq}$ is the input to the error dynamical model. 

It is worthy mentioning that the dynamical model (\ref{dynamic}) is built on the assumption that both the number of effective coil turns $N$ and the effective magnetic pole area $A$ are constant, which strictly speaking is not sensible.  Moreover, $m$ is almost impossible to obtain exactly in practice. The disturbance $f$ includes the wind resistance and coefficients $A$, $N$ and $\kappa$ vary with the commute of the mlsMT. For example, the effective magnetic pole area $A$ varies due to the motion of the vehicle. And to derive (\ref{dynamic}), it further uses 
\begin{equation}
\phi = \frac{F}{R}= \frac{Ni(t)}{2x(t)/(\mu_0 A)}
\end{equation}
as the magnetic flux with the ignorance of its leakage and magnetoresistance of the electromagnet core and the rail. Overall, the dynamical model (\ref{dynamic}) is a very simplified version and contains many unmeasurable coefficients. 

\begin{figure}[t]
\centering
\subfigure{
\includegraphics[width=42mm]{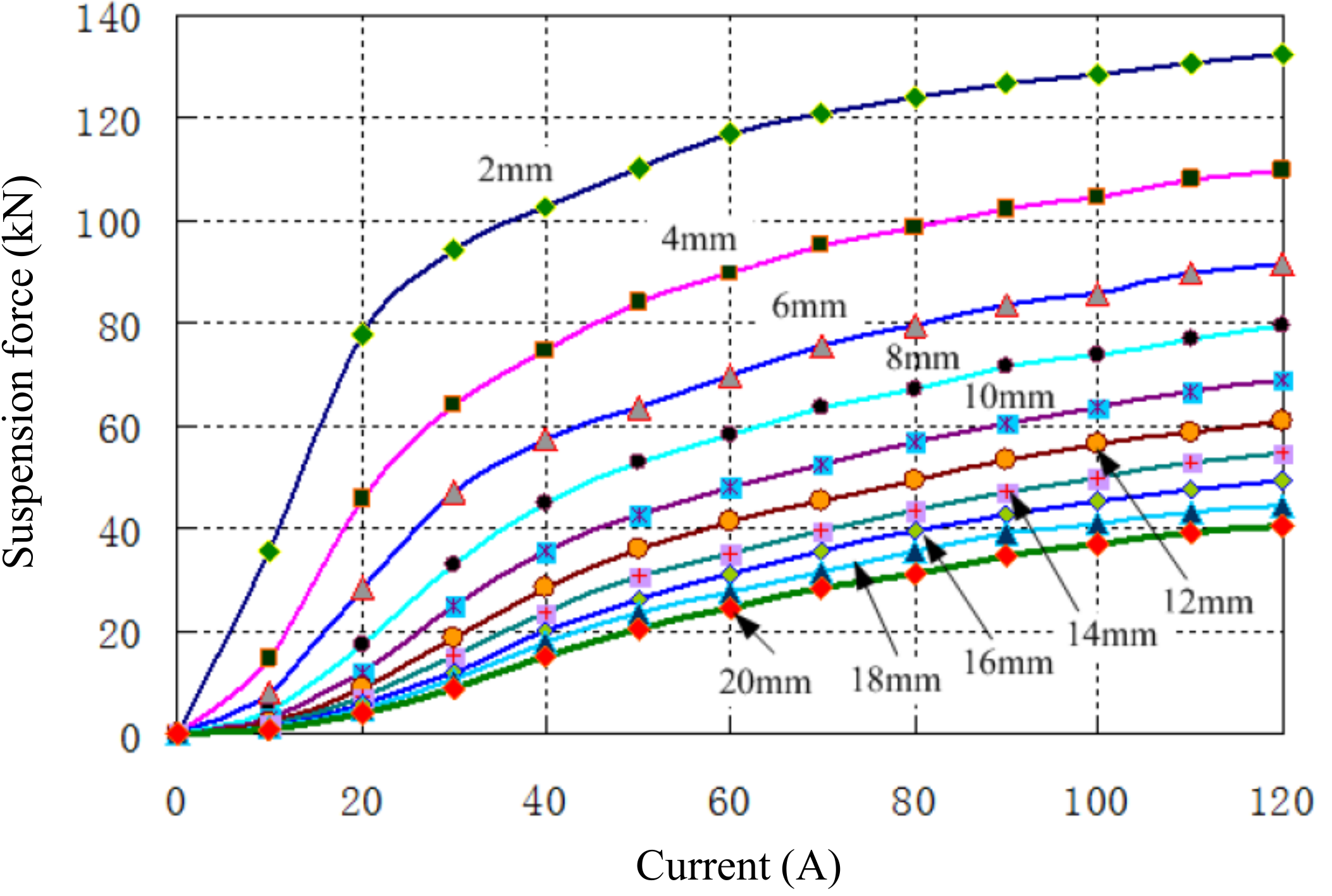}}
\subfigure{
\includegraphics[width=42mm]{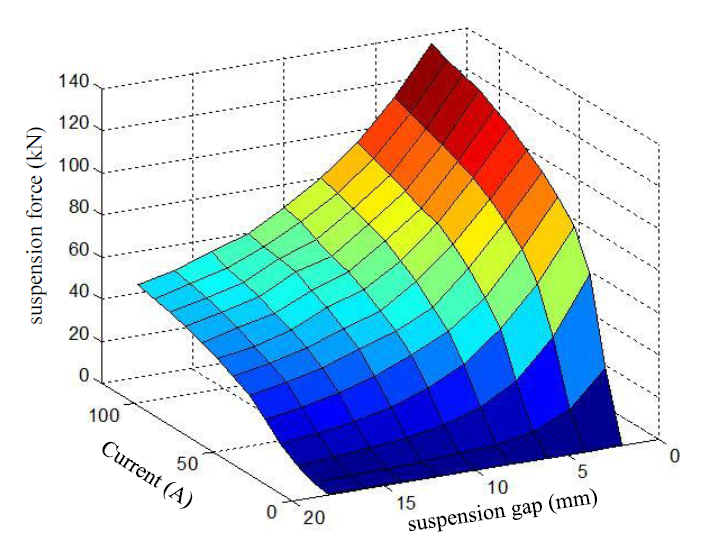}}
\caption{The suspension force at different currents and air gaps.}
\label{force}
\end{figure}

Thus, the model-based controllers relying on (\ref{dynamic}) may not be very reliable, although most of the model-based controllers in Section \ref{sec_intro} do so. For instance, a linearization approach has been adopt to control the dynamical model in (\ref{dynamic}), which is in fact not sensible. To illustrate this, experiments in the Changsha's mlsMT lab has been conducted to analyze the electromagnetic force at each pair of current $i(t)$ and air gap $x(t)$, see Fig. \ref{force}. The gradient of the suspension force varies rapidly with the current and air gap, which indicates that the nonlinear term cannot be neglected. Thus the control performance based on the linearized model is expected to deteriorate rapidly with an increasing deviation from nominal operating points.

\subsection{The Objective of This Work}
The purpose of this work is to design sample-based reinforcement learning (RL) algorithms to learn an effective controller to regulate the air gap between the electromagnet and track to a desired level. The advantages of the RL framework to address the suspension regulation problem at least include: (a) The dynamical model of the magnetic levitation system is difficult to obtain in practice, while our controller is built based on samples, instead directly on the knowledge of the dynamical model; and (b) the uncertain system dynamics and external disturbances can be regarded as the components of the external environment, whose details are not required by the RL algorithms. Moreover, the mlsMT always run in the same rail, implying that the training data is recyclable, which suits RL algorithms well.

\section{The MDP for Suspension Regulation}\label{sec_mdp}
In this section, we reformulate the suspension regulation problem as a MDP with unknown transition probabilities due to unknown dynamics of the magnetic levitation system.
\subsection{Markov Decision Process}\label{AA}
A reinforcement learning task that satisfies the Markov property is called a Markov decision process, or MDP, where the Markov property means that the current state of an agent contains all the relevant information. The MDP is described by four components: $(a)$ a state space $\mathcal{S}$; $(b)$ an action space $\mathcal{A}$; $(c)$ an one-step cost function $c(\bm{s}, \bm{a}):\mathcal{S}\times \mathcal{A} \to \mathbb{R}$; and $(d)$ a stationary one-step transition probability $p(\boldsymbol{s}_k|\boldsymbol{s}_1,\boldsymbol{a}_1,...,\boldsymbol{s}_{k-1},\boldsymbol{a}_{k-1})$. The Markov property ensures that the current state only depends on the last state and relative action, i.e.,
\begin{equation}
p(\boldsymbol{s}_k|\boldsymbol{s}_1,\boldsymbol{a}_1,...,\boldsymbol{s}_{k-1},\boldsymbol{a}_{k-1})
    =p(\boldsymbol{s}_k|\boldsymbol{s}_{k-1},\boldsymbol{a}_{k-1}).
\end{equation}

\begin{figure}[t]
\centerline{\includegraphics[width=70mm]{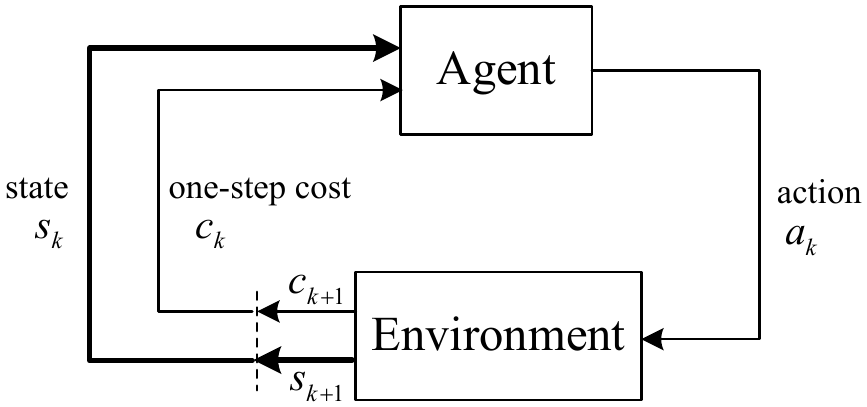}}
\caption{The agent-environment interaction in a MDP \cite{sutton2011reinforcement}.}
\label{mdp}
\end{figure}

The MDP describes how an agent interacts under the environment. The agent takes an action $\bm{a}_k$ under the current state $\bm{s}_k$, and then observes the next state $\bm{s}_{k+1}$. Meanwhile, the agent receives a one-step cost $c_k=c(\bm{s}_k, \bm{a}_k)$, as illustrated in Fig. \ref{mdp}.
A policy is a mapping from the state space $\mathcal{S}$ to the action space $\mathcal{A}$, and can be defined as a function $\pi:\mathcal{S}\to\mathcal{A}$.  The purpose of RL algorithms is to find an optimal policy $\pi$ that minimizes the long-term cumulative cost. Specifically, it is defined as the sum of all discounted one-step cost. Then an optimal policy is obtained by solving the following optimization
\begin{equation}
    \min\limits_{\pi \in \mathcal{P}}J(\pi)= \min\limits_{\pi \in\mathcal{P}}
    E
    \left[
    \sum_{k=1}^{K}\gamma^{k-1}c_k|\pi
    \right]
\end{equation}
where $\mathcal{P}$ is the policy space,  $\gamma$ with $0<\gamma<1$ is a discounted rate that measures the decay of cost over time, $K$ denotes the horizon of the problem, and the expectation is taken over the stochastic process.

It is essential to properly define the four components of MDPs for the suspension regulation problem. In this work, the action space is the control input of the magnetic levitation system, i.e., the voltage of electromagnet.  The transition probability is unknown and needs to be learnt from environment. Thus we mainly focus on the design of the state space and the one-step cost. Note that we formulate the suspension regulation problem as a continuous-state, continuous-action MDP.
\subsection{The MDP for Suspension Regulation}
The purpose of the suspension regulation problem is to ensure the air gap stable between electromagnet and the track with minimum energy consumption. That is, both the regulation error  $x_k-x_{eq}$ and its derivative $\dot{x}_k$ should be well controlled to be zero. Since the input voltage $u_k$ measures the consumption of energy, the one-step cost of the MDP at the step $k$ is defined as follows
\begin{equation} \label{cost}
    c(\boldsymbol{s}_k, u_k)=\rho_1(x_k-x_{eq})^2+\rho_2\dot{x}_k^2+\rho_3u_k^2,
\end{equation}
where $\rho_i, i\in\{1,2,3\}$, are positive constants and $x_{eq}$ is the desired air gap in \eqref{state_sys}.

There are two essential principles for the design of the state. Since the policy is a mapping from the state space to the action space,  the information contained in the state should be sufficient, which ensures that the system is completely controllable with the action as the input. For example, if the state only contains the air gap $x_k$,  the system in the state $x_k$ may have derivative of two possible cases, i.e., $\dot{x}_k>0$ and $\dot{x}_k\leq 0$. If the action is merely determined by the state $x_k$, the system may not be stabilized in any case. Thus, the state of MDP needs to fully describe the system. On the other hand, the state cannot be too redundant due to the fact that a high dimension state greatly increases the computational cost and the difficulty of solving the MDP. For the suspension regulation problem of this work, we observe that the state variables 
\begin{equation}
    \bm{s}_k = [x_k-x_{eq}, \dot{x}_k, i_k-i_{eq}]^\mathrm{T}.
\label{state}
\end{equation} well fulfill the above principles, and are selected as the state of MDP.

\section{Solving the MDP via DDPG}\label{sec_ddpg}
\label{sec_solve}
In this section, we design RL algorithms with Deep Deterministic Policy Gradient (DDPG) to solve the MDP for the suspension regulation.

\subsection{Dynamic Programming}
Dynamic programming~\cite{sutton2011reinforcement} is a backward algorithm to find an optimal policy with a full MDP model. It provides basic ideas that can be applied to more general cases. Here we introduce two value functions to evaluate the quality of a policy ${\pi}$. The first value function is a long-term cost from the current state
\begin{equation}
    \mathit{v_\pi}(\bm{s})=E_{\pi}
    \left[
    \sum_{k=1}^K\gamma^{k-1}c_{k}| \bm s_1=\bm{s}
    \right].
\end{equation}
 The other is the Q-value function defined as the value of taking action $\bm a$ in state $\bm s$ under a policy $\pi$
\begin{equation}
q_\pi(\bm s, \bm a)=E_\pi
\left[
\sum_{k=1}^K\gamma^{k-1}c_{k}| \bm s_1=\bm{s},\bm a_1=\bm a
\right].
\end{equation}

A fundamental property for the value function $v_\pi(\cdot)$ in RL and dynamic programming is that they satisfy the well-known {\em Bellman equation}, i.e., 
\begin{equation}
v_{\pi}(\bm{s})=\sum_{\bm a}\pi(\bm a | \bm s)
    \sum_{\bm{s'}} p(\bm s' | \bm s, \bm a)
    [
        r(\bm s, \bm a, \bm s')+\gamma  v_\pi(\bm s')
    ].
\end{equation}

In dynamic programming, an optimal policy can be obtained through the iteration between policy evaluation and policy improvement.  To evaluate a policy $\pi$, Bellman equation provides a recursive way to compute the value function with an initial value function $\mathit v_\pi^0(\bm s)$, i.e., 
\begin{equation}
\mathit{v_{\pi}^{j+1}}(\bm s)  = 
\sum_{\bm s'}p(\bm s' | \bm s, \bm a)
[
r(\bm s, \bm a, \bm s')+\gamma \mathit v_{\pi}^{j}(\bm s')
], j\in\mathbb{N}.
\end{equation}

The iteration can be computed until the convergence or for any fixed number of steps. After the policy evaluation, we do policy improvement to obtain an improved policy based on the estimated value function by a greedy minimization, i.e.,
\begin{equation}
    \pi'(\bm s) = \mathop{\arg\min}\limits_{\bm a \in \mathcal{A}(\bm s)}
\sum_{\bm s'}p(\bm s' | \bm s, \bm a)
[
    r(\bm s, \bm a, \bm s')+\gamma \mathit v_{\pi}(\bm s')
].
\end{equation}

Then, the policy is evaluated and improved alternatively until the policy converges to an optimal one.
\subsection{Neural Network Approximators}

\begin{figure}[t]
\centering
\subfigure[The critic network]{
\centerline{\includegraphics[width=70mm]{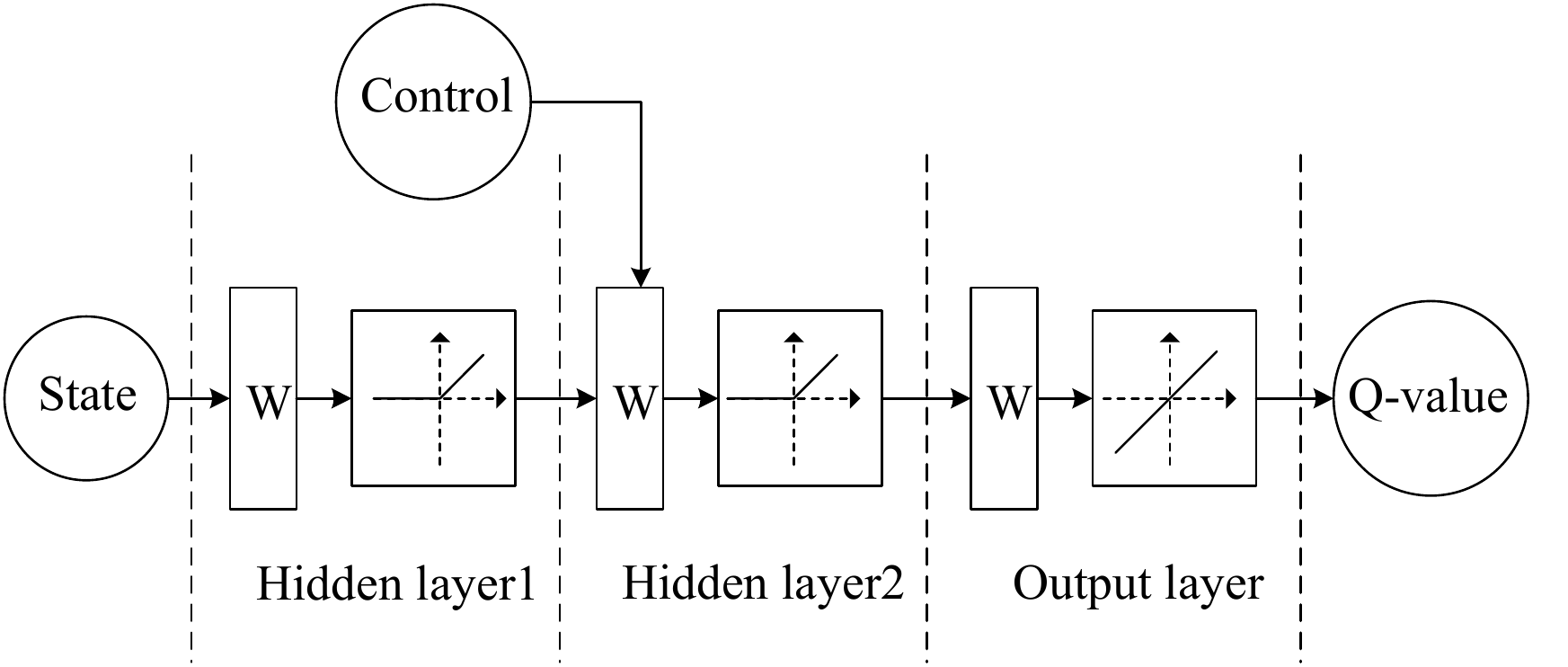}}
}
\subfigure[The actor network]{
\centerline{\includegraphics[width=70mm]{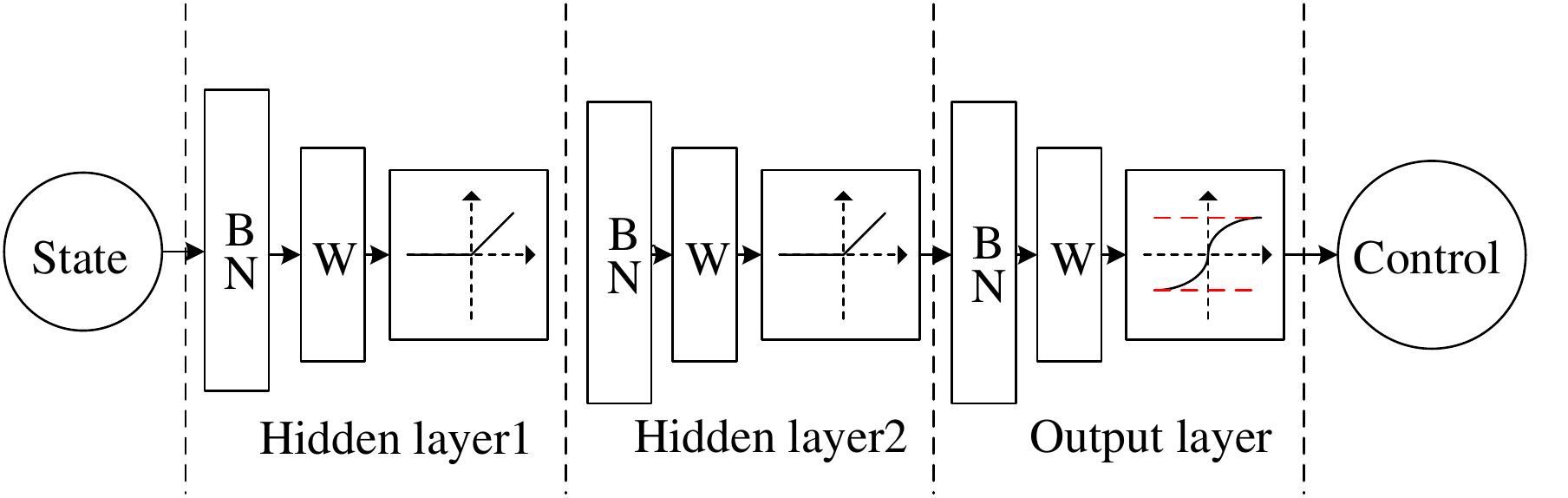}}
}
\caption{Structures of two neural networks.}
\label{networks}
\end{figure}

For the discrete-state and discrete-action MDP, dynamic programming saves the Q-value $q_\pi(\bm s, \bm a)$ and the policy $\pi(\bm s)$ in a lookup table. Unfortunately, tabular methods cannot be applied to the MDP with continuous state space and action space. In the suspension regulation problem in Section \ref{sec_problem}, the air gap and its derivative, the current and input voltage are all continuous. In this case, the Q-value and policy need to be approximated with parametrized functions, respectively. Due to strong nonlinear couplings of the magnetic levitation system, we propose the critic neural network $q(\bm s, u|\omega)$ and the actor neural network $\mu(\bm s| \theta)$,where $\omega$ and $\theta$ are parameters to be trained,  as approximators to the Q-value function and policy function respectively.

\begin{figure}[t]
	\centerline{\includegraphics[width=60mm]{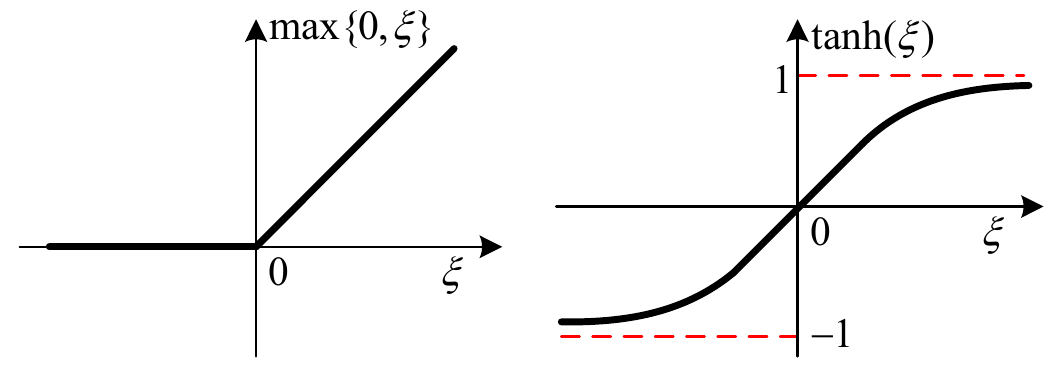}}
	\caption{Curves of the activation functions. The Relu function $\max \{0, \xi\}$ is on the left, while the tanh function $tanh(\xi) = \frac{e^{\xi}-e^{-\xi}}{e^{\xi}+e^{-\xi}}$ is on the right.}
	\label{relu}
\end{figure}

Inspired by \cite{lillicrap2017continuous}, a critic network for approximating the Q-value function is designed to have three layers with the state $\bm s$ and the control $u$ as network inputs, and the control to the magnetic levitation system is also added in the second layer. See Fig. \ref{networks} for an illustration. The activation function of two hidden layers is selected as a Relu function which can accelerate the convergence and avoid gradient vanishing~\cite{hochreiter2001gradient}.  The output layer adopts a linear function as the activation function to generate a scalar Q-value. The Relu and tanh functions are illustrated in Fig.~\ref{relu}.

In contrast, the actor network for approximating the policy function has three layers, each of which also consists of a batch normalization layer~\cite{ioffe2015batch}. The network input is the state $\bm s$ and it generates control as the output. Due to physical constraints of mlsMTs, the control voltage is limited to a given range. Thus, we adopt a tanh unit as the activation function of output layer to generate values in $[-1, 1]$ and then scale them to satisfy the constraint.  In the actor network,   batch normalization is mainly used to eliminate the value difference among the components of states and accelerate the convergence of training neural networks. Batch normalization is not necessary for the critic network since the absolute value of the Q-function is immaterial and only its relative value is used for the improvement of the policy.

\subsection{Temporal Difference and Deterministic Policy Gradient}
Since the transition probabilities $p(\bm s' | \bm s, \bm a)$ are unknown in the magnetic levitation system, Temporal Difference (TD) methods ~\cite{sutton2011reinforcement} are adopt to learn directly from raw experiences. The differences between dynamic programming and TD methods are primarily in their approaches to do policy evaluation.

In the sequel, we substitute the action $\bm a$ in the MDP with the input voltage $u$ in the suspension regulation problem. Suppose that we observe a triple $(\bm s_{k}, u_{k}, \bm s_{k+1})$ at time $k$, then TD methods update Q-value function as follows
\begin{equation}
\begin{aligned}
q(\bm s_k, u_k) \gets &q(\bm s_k, u_k) \\
&+\alpha[c_{k}(\bm s_k, u_k)+ \gamma\min\limits_{u}q(\bm s_{k+1}, u)-q(\bm s_k, u_k)]
\end{aligned}
\label{Q-value}
\end{equation}
 where $\alpha$ is a learning rate.

Here we use the Q-value function for the policy improvement. We approximate the Q-value as a continuous function  $q(\bm s, u| \omega)$ with a parameter vector $\omega$, and update it as follows:
\begin{eqnarray}
\begin{aligned}
\delta_k= & \ c(\bm s_k, u_k) \\
& +\gamma q(\bm s_{k+1}, \mu(\bm s_{k+1}| \theta)|\omega)
-q(\bm s_k, u_k | \omega) \\
\omega_{k+1}= & \ \omega_{k}+\alpha \delta_k \nabla_\omega q(\bm s_k, u_k | \omega)
\end{aligned}
\end{eqnarray}
where $\delta_k$ is called TD error and $\mu(\bm s_{k+1}| \theta)$ is the approximated policy function. This forms the policy evaluation.

For the policy improvement, we adopt the deterministic policy gradient algorithm~\cite{silver2014deterministic} and use the gradient descent method to minimize the long-term cost function $J(\theta)$ under a given policy $\mu(\bm s | \theta)$, i.e.,

\begin{equation}
\min\limits_{\theta}J(\theta)= \min\limits_{\theta}E
\left[
\sum_{k=1}^{K}\gamma^{k-1}c_k|\theta
\right].
\end{equation}

However, due to the unknown transition probability, the gradient of $J(\theta)$ cannot be computed directly. The deterministic policy gradient (DPG) algorithm solves the minimization with an approximation of the gradient~\cite{silver2014deterministic}.

In the DPG algorithm, the parameter vector $\theta$ is updated along the negative gradient of the long-term cost function $J(\theta)$,
\begin{equation}
\theta_{k+1}\doteq \theta_k - \alpha \widehat{\nabla_{\theta}J(\theta_k)}, 
\end{equation}
where $\widehat{\nabla_{\theta}J(\theta_k)}$ is an unbiased estimate of the true gradient of $J(\theta)$ evaluated at $\theta=\theta_k$. With the approximated Q-value, the DPG algorithm leads to the form of $\widehat{\nabla_{\theta}J(\theta)}$ as follows

\begin{equation}
\widehat{\nabla_{\theta}J(\theta)} \approx \\
\frac{1}{B}\sum_{i=1}^B\nabla_{\theta}\mu(\bm s_i | \theta) \nabla_{u_i}q(\bm s_i, u_i|\omega)
\end{equation}
where $B$ is the size of batch.

\begin{algorithm}[t]
	\caption{The DDPG for  suspension regulation}
	\label{alg:DDPG}
	\begin{algorithmic}[1]
		\Require
		Target air gap $x_{eq}$, number of episodes $M$, number of steps for each episode $K$, batch size $B$, learning rates for two networks $\alpha_\omega$ and $\alpha_\theta$.
		\Ensure
		suspension regulation policy $u = \mu(\bm s| \theta)$.
		\State Initialize critic networks $q(\bm s, u|\omega)$ and the actor network $\mu(\bm s|\theta)$. \\
		Initialize target networks $\omega^{\prime} \leftarrow \omega,  \theta^{\prime} \leftarrow \theta$. \\
		Initialize the experience replay cache R.
		\For{$m=1 \ to \ M$}
		\State Reset the initial state $\bm s_0$
		\For{$k=0 \ to \ K$}
		\State
		Generate a control input $u_k=\mu(\bm s_k|\theta) + \triangle u_k$  \phantom{..........} with current actor network and exploration noise  \phantom{...........}
		$\triangle u_k \sim \mathcal{N}(0, \tilde{\sigma})$.
		\State Execute control $u_k$ and obtain $\bm s_{k+1}$. Calculate one-\phantom{............}step cost $c_k$ according to (\ref{cost}).
		\State Push experience $(\bm s_k, u_k, c_{k+1}, \bm s_{k+1})$ into $R$.
		\State Randomly sample a minibatch $$\{(\bm s_j, u_j, c_{j+1}, \bm s_{j+1}) | 1 \leq j \leq B\}$$ \phantom{...........} from cache R.
		\State Update the  critic network:
		\phantom{...........}
		\begin{equation}
		~~~~~\begin{aligned} 
		\label{critic_ddpg}
		y_j &= c_{j+1} + \gamma q(\bm s_{j+1}, \mu(\bm s_{j+1}|\theta^{\prime})|\omega^{\prime}) \\
		\delta_j &=  y_j - q(\bm s_j, u_j|\omega_{i}) \\
		\omega &\leftarrow \omega + \frac{1}{B}\alpha_\omega \sum_{j=1}^B \delta_j \nabla_{\bm w}q(\bm s_j, u_j|\omega)
		\end{aligned}
		\end{equation}

		\State Compute $\nabla_{ u_j}q(\bm s_j, u_j|\omega)$ for each sample.
		\State Update the actor network:
		\begin{equation}
		\phantom{...........}
		\theta \leftarrow \theta - \frac{1}{B}\alpha_{\bm_{\theta}} \sum_{j=1}^B\nabla_{\theta}\mu(\bm s_j | \theta) \nabla_{ u_j}q(\bm s_j, u_j|\omega)
		\label{actor_ddpg}
		\end{equation}
		\phantom{...........} Update the target networks: 
		\begin{equation}
		\begin{array}{l}{\omega^{\prime} \leftarrow \tau \omega+(1-\tau) \omega^{\prime}} \\ {\theta^{\prime} \leftarrow \tau \theta+(1-\tau) \theta^{\prime}}\end{array}
		\end{equation}
		\EndFor
		\EndFor
	\end{algorithmic}
\end{algorithm}
 
Till now, we have defined the Q-value function $q(\bm s, u|\omega)$ and policy function $\mu(\bm s| \theta)$ and given their update rules. The RL algorithm is based on an actor-critic framework where the critic (Q-value) approximator is used to learn the optimal costs and the actor (policy) approximator is used to learn optimal control policies. 
One of the dominant actor-critic methods is Deep Deterministic Policy Gradient algorithm (DDPG), which is first proposed for the continuous control problems in \cite{lillicrap2015continuous}. In DDPG, the critic network is updated using the TD method with a secondary frozen target network $q(\boldsymbol{s}, u | \omega^{\prime})$ to maintain a fixed objective $y_k$ over multiple updates, i.e., 
\begin{equation}
y_k=c_{k}(\bm s_k, u_k) + \gamma q(\bm s_{k+1}, u_{k+1} | \omega^{\prime}), \ \ u_{k+1}= \mu(\bm s_{k+1}| \theta^{\prime}),
\end{equation}
where the actions $u_{k+1}$ are selected from a target actor network $\mu(\bm s| \theta^{\prime})$. The weights of the target networks are then updated by a proportion $\tau$ at each time step 
\begin{equation}
\begin{aligned}
\omega^{\prime}& \leftarrow \tau \omega+(1-\tau) \omega^{\prime}\\
\theta^{\prime}& \leftarrow \tau \theta+(1-\tau) \theta^{\prime}.
\end{aligned}
\end{equation}

Using a frozen target network to provide the target value for the network update improves the stability of the convergence, and thus accelerates training.

\begin{figure}[t]
	\centerline{\includegraphics[width=70mm]{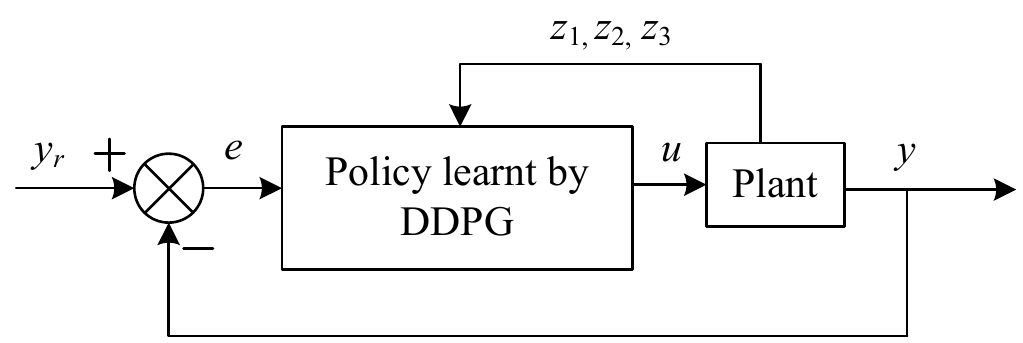}}
	\caption{Structure diagram of the magnetic levitation system with the controller provided by the DDPG algorithm.}
	\label{dpg}
\end{figure}
\subsection{Experience Replay}
Note that the approximate gradient is calculated by a transition sequence, which is inconvenient for the online suspension regulation problem. Actually, a batch learning scheme is performed to calculate the approximate gradient with  past experiences.
We now discuss how to obtain the samples of trajectories and use them to train the actor network and the critic network. Instead of designing the controller based on an accurate model, our RL algorithm learns the optimal policies from  trajectories of the magnetic levitation system. However, due to limited memory storage and expensive cost for a trial, the volume of sampled data may be insufficient. Therefore, it is essential to learn an optimal controller with a moderate size of  sampled data.

The above TD methods adjust the network with the most recent experience only once and then dump it. We apply a batch learning scheme called {\em experience replay}~\cite{lin1992self} to improve the data efficiency. In the MDP, the agent executes an action $\bm a$ at a state $\bm s$ and observe a new state $\bm s'$ and an one-step cost $c$. An {\em experience} (or a sample) is defined as a quadruple $(\bm s, \bm a, \bm s', c)$. With the exploration of the agent, the size of experiences grows up which are added to a cache for the experience replay. Instead of using the new experience, a batch of experiences is randomly sampled from the cache for updating the two neural networks. By experience replay, the learning agent repeatedly accesses past experiences as if it experienced over and again what it had experienced before. 

The motivation of experience replay is also to remove the correlation of the sampled data and address the problem of the non-stationary distribution of experiences. Thus, it improves the data efficiency for the recycling of experiences, which is relatively suitable for the suspension regulation problem of mlsMTs.

From the control point of view, the policy learnt by the DDPG algorithm can be considered as a full state feedback controller with the state (\ref{state}) as the input and the action as the output, where the state $\bm s_{k}=[x_k-x_{eq}, \dot{x}_k, i_k-i_{eq}]^\mathrm{T}$ can be measured directly in practice. Since the state (\ref{state}) in the MDP is consistent with the state (\ref{state_sys}) of the control system, we use $\bm z = [z_1, z_2, z_3]^\mathrm{T}$ to denote the state of the control system.  Fig.~\ref{dpg} illustrates the overview of the magnetic levitation system, where $y$ and $y_r$ denotes the output and reference input. The details of the DDPG algorithm for the suspension regulation is given in Algorithm \ref{alg:DDPG}.

\section{Solving the MDP via TD3}\label{sec_td3}
In Section \ref{sec_solve}, we describe the RL algorithm structure with two neural network approximators, i.e., the critic network $q(\bm s, u|\omega)$ and the actor network $\mu(\bm s| \theta)$. While the approximation inevitably introduces the function approximation error, it may reduce the learning speed~\cite{fujimoto2018addressing}. To improve the stability of convergence, we further propose the twin delayed deep deterministic policy gradient (TD3, \cite{fujimoto2018addressing}) method for the suspension regulation of mlsMTs.

\begin{algorithm}[t!]
	\caption{The TD3 for the suspension regulation}
	\label{alg:Framwork}
	\begin{algorithmic}[1]
		\Require
		Target air gap $x_{eq}$, number of episodes $M$, number of steps for each episode $K$, batch size $B$, learning rates for two networks $\alpha_\omega$ and $\alpha_\theta$.
		\Ensure
		suspension regulation policy $u = \mu(\bm s| \theta)$.
		\State Initialize two critic networks $q(\bm s, u|\omega_{1}), q(\bm s, u|\omega_{2})$ and the actor network $\mu(\bm s|\theta)$. \\
		Initialize target networks $\omega_{1}^{\prime} \leftarrow \omega_{1}, \omega_{2}^{\prime} \leftarrow \omega_{2}, \theta^{\prime} \leftarrow \theta$. \\
		Initialize the experience replay cache R.
		\For{$m=1 \ to \ M$}
		\State Reset the initial state $\bm s_0$
		\For{$k=0 \ to \ K$}
		\State
		Generate a control input $u_k=\mu(\bm s_k|\theta) + \triangle u_k$  \phantom{..........} with current actor network and exploration noise  \phantom{...........}
		$\triangle u_k \sim \mathcal{N}(0, \tilde{\sigma})$.
		\State Execute control $u_k$ and obtain $\bm s_{k+1}$. Calculate one-\phantom{............}step cost $c_k$ according to (\ref{cost}).
		\State Push experience $(\bm s_k, u_k, c_{k+1}, \bm s_{k+1})$ into $R$.
		\State Randomly sample a minibatch $$\{(\bm s_j, u_j, c_{j+1}, \bm s_{j+1}) | 1 \leq j \leq B\}$$ \phantom{...........} from cache R.
		\State Update the pair of critic networks:
		\phantom{...........}
		\begin{equation}
		~~~~~~~~\begin{aligned} 
		y_j &= c_{j+1} + \gamma \max _{i=1,2} q(\bm s_{j+1}, \mu(\bm s_{j+1}|\theta^{\prime})+\epsilon|\omega^{\prime}_{i}) \\
		\delta_j &=  y_j - q(\bm s_j, u_j|\omega_{i}) \\
		\omega_{i} &\leftarrow \omega_{i} + \frac{1}{B}\alpha_\omega \sum_{j=1}^B \delta_j \nabla_{\bm w}q(\bm s_j, u_j|\omega_{i}), i \in\{1,2\}
		\end{aligned}
		\label{critic_td3}
		\end{equation}
		\phantom{............}where $\epsilon  \sim \operatorname{clip}(\mathcal{N}(0, \sigma),-l, l)$.
		\If{$k \ mod \ d$}
		\State Compute $\nabla_{ u_j}q(\bm s_j, u_j|\omega_{1})$ for each sample.
		\State Update the actor network:
		\begin{equation}
		\phantom{................}
		\theta \leftarrow \theta - \frac{1}{B}\alpha_{\bm_{\theta}} \sum_{j=1}^B\nabla_{\theta}\mu(\bm s_j | \theta) \nabla_{ u_j}q(\bm s_j, u_j|\omega_{1})
		\label{actor_td3}
		\end{equation}
		\phantom{................} Update the target networks: 
		\begin{equation}
		\begin{array}{l}{\omega_{i}^{\prime} \leftarrow \tau \omega_{i}+(1-\tau) \omega_{i}^{\prime}} \\ {\theta^{\prime} \leftarrow \tau \theta+(1-\tau) \theta^{\prime}}\end{array}
		\end{equation}
		\EndIf
		\EndFor
		\EndFor
		
	\end{algorithmic}
\end{algorithm}

The actor-critic methods in Section \ref{sec_solve} use the critic neural network to estimate the Q-value function for the policy evaluation. Note that the Q-value function is always underestimated for the minimization of a noisy Q-value function in the update. By (\ref{Q-value}), this underestimated bias is further exaggerated by the TD methods in which the error is accumulated throughout the iterative Q-value evaluation. To reduce the bias, a clipped double Q-learning algorithm is proposed, in which the Q-value is estimated as the maximum between two independent critic neural networks, e.g., $q(\bm s, u|\omega_{1})$ and $q(\bm s, u|\omega_{2})$. Then, the target of updating two critic networks is
\begin{equation}
y_{k}= c_{k}(\bm s_k, u_k) + \gamma \max _{i=1,2} q(\bm s_{k+1}, u_{k+1} | \omega_{i}^{\prime}), 
\end{equation}
where $u_{k+1}= \mu(\bm s_{k+1}| \theta^{\prime})$.
With the clipped double Q-learning, the underestimate of the Q-value is eliminated~\cite{fujimoto2018addressing}. Although this update rule may also lead to the underestimate bias, this is more acceptable since the actions with high value will not be propagated through TD methods.

The high variance of the underestimation bias is also a significant factor that provides a noisy gradient for the policy update, which reduces the learning speed. The Q-value estimates diverge due to underestimation when the policy is poor, and the policy becomes poor when the Q-value is estimated inaccurately.
Therefore the failure of the actor-critic methods occurs in the interplay between the actor and critic updates. To address this problem, the actor network should be updated at a lower frequency than the critic network to provide a stable policy for the Q-value estimate. This ensures that the TD error becomes small enough before the next update of the policy. 

The variance can be further reduced through the regularization for value learning and policy smoothing. An intuitive idea is that similar actions should have similar value, which can be enforced though adding a small noise $\epsilon$ into the target update of the value in the training procedure, that is
\begin{equation}
\begin{aligned}
&y_k=c_{k}(\bm s_k, u_k)+\gamma \mathbb{E}_{\epsilon}\left[q(\bm{s}_{k+1}, u_{k+1}+ \epsilon | \omega^{\prime})\right], \\
 &u_{k+1} = \mu(\bm s_{k+1}| \theta^{\prime}).
\end{aligned}
\end{equation}

In implementation, we can approximate the expectation by sampling noisy actions of the target policy and averaging over mini-batches. The modified target update becomes:
\begin{equation}
\begin{aligned} y &=c_{k}(\bm s_k, u_k)+\gamma q(\bm{s}_{k+1}, u_{k+1}+ \epsilon | \omega^{\prime}), \\  \epsilon & \sim \operatorname{clip}(\mathcal{N}(0, \sigma),-l, l), \end{aligned}
\end{equation}
where the added noise $\epsilon$ is sampled from a Gaussian distribution $\mathcal{N}(0, \sigma)$ with $\sigma$ as its variance, and then clipped to keep the action around the original output of the policy network.

Then we propose the Twin Delayed Deep Deterministic policy gradient algorithm (TD3) for the suspension regulation which is built on the DDPG. Particularly, the simulation shows that the TD3 improves the stability and performance of the DDPG algorithm.

To sum up, we propose a TD3 algorithm for the suspension regulation with neural networks as approximators. This algorithm is a sample-based method without the knowledge of the dynamical model of the magnetic levitation system, and its performance is comparable to some model-based control methods under the exact dynamical model. A detailed procedure of the TD3 algorithm is given in Algorithm~\ref{alg:Framwork}.

\section{Algorithm Implementation}\label{sec_alg}
In this section, we discuss several important details in the algorithm implementation and an initialization method that can significantly improve the performance of RL algorithm for the suspension regulation.
\subsection{Implementation Details}
The simulation of the RL algorithms includes four main components, i.e., the setting of the simulation model, the initialization, the interaction and the training of networks. The details are provided as follows.
\begin{enumerate}
\renewcommand{\labelenumi}{\rm(\alph{enumi})} 
\item The construction of our RL based controller does not require the direct knowledge of the dynamical model, but a simulation model is needed to simulate the physical magnetic levitation system of the mlsMT. For example, the nonlinear model in (\ref{dynamic}) can be used as the simulator, into which both the noise and the external disturbance are added. Notice that the simulation model is only used to generate samples, i.e., a quadruple $(\bm s, u, \bm s', c)$, from which the RL algorithm learns a feedback controller.

\item The initialization mainly refers to the construction of experience replay cache $R$ and the neural networks, i.e., a pair of critic networks $q(\bm s, \bm a|\omega_{i})$ and the actor network $\bm u = \mu(\bm s| \theta)$. The cache $R$ can either be empty when all needed experiences come from the online interaction between the agent and the environment, or filled with initial data, which can be obtained from previous running of the program, or even the experiences generated on other controllers. 

\item The agent is set to an initial state at the beginning of each episode, then executes the action according to the actor network. The interaction between the agent and the environment returns an experience, $(\bm s, u, \bm s', c)$, which is then added into the experience replay cache $R$. A minibatch of experiences can be extracted to train the neural networks as soon as its total number reaches the batch size, e.g., it is usually set to 32 or 64. To ensure a stable training, in our experiemnt we use a purely exploratory policy for the first 10000 time steps then begin to train the networks.

\item The training of the critic network involves the computation of Q-value error $\delta_j$ and the gradient $\nabla_{\omega}q(\bm s_j, u_j|\omega)$ according to (\ref{critic_ddpg}) and (\ref{critic_td3}). This is easy to implement using the back propagation algorithms. The update of the actor networks (\ref{actor_ddpg}) and (\ref{actor_td3}) requires the gradient of the network, which can be approximated through the deterministic policy gradient. In TD3, a delay update rule is adopted to provide a stable policy for the Q-value estimation. 
\end{enumerate}

\subsection{Learning From Existing Controllers}
Sample-based methods can learn optimal controllers for a wide range of the continuous control tasks, but typically require a very large number of samples to achieve a good performance~\cite{nagabandi2018neural}. In the experiments the RL algorithm may fail to converge when the sampled trajectories are poor at the beginning episodes. This means that some state-action pairs are rarely explored by the initial policy, which results in an unbalanced distribution, thus reducing the learning speed. To achieve enough exploration, the agent has to generate a large number of samples, while most of them have little contribution to the training of the networks. Generally the improvement of the policy largely relies on the low-cost samples where a good action is exerted at the state.

To address it, an intuitive idea is that we can combine the samples generated by the RL algorithms with the trajectories sampled from other controllers. This method is subsumed within the well-known class of {\em transfer learning} \cite{pan2009survey,taylor2009transfer}, which seeks to transfer external knowledge into a new task to help learning. For example, we can firstly design a coarse PID controller, and then generate enough control trajectories to initialize the experience replay buffer. Thus the initial distribution of samples is well balanced and the agent learns both from good and bad states. In implementation, we analyze the real samples provided by CRRC obtained using PID controller and manually select the state-action pairs to ensure a satisfactory distribution, then we initialize the replay cache with these samples. We illustrate with simulations that the training is accelerated notably and the final performance is improved. 

\section{Experiment Study}\label{sec_exp}
In this section, we compare the performance of our controller learnt by RL algorithms with the model-based controller by simulations. We adopt the PID controller and the linear quadratic gaussian integral (LQI) controller as comparisons. To this end, the nonlinear dynamical model (\ref{dynamic}) is linearized at the equilibrium point and then two controllers are designed based on its linearized model. We describe the experiment setting of the RL algorithms and simulate the state feedback controller learnt by DDPG and TD3, respectively, to show the distinguished improvement. We also test our RL framework on the real dataset sampled from the trial of mlsMTs in Changsha, China.

\subsection{Proportional-Integral-Derivative Control}
We first present the values of dynamic coefficients in (\ref{dynamic}) in Table \ref{table}.

\begin{table}[t]
\caption{DYNAMIC PARAMETERS}
\label{table}
\begin{center}
\begin{tabular}{|cc|cc|}
\hline
Coefficients & Value & Coefficients & Value\\
\hline
$m/kg$ & $700$ & $N$ & $450$ \\

$A/m^2$ & $0.024$ & $R/ \Omega$ & $1.2$  \\

$i_{eq}/A$ & $17.0$ & $x_{eq}/m$ & $0.008$  \\

$\mu_0/(H\cdot m^{-1})$ & $4\pi \times 10^{-7} $& $g/(m \cdot s^{-2})$ & 9.8 \\

\hline
\end{tabular}
\label{tab1}
\end{center}
\end{table}

Assuming that the magnetic levitation system are at the equilibrium point, the air gap is $x_{eq}$, the derivative of the air gap is $0$, and the current in the coil is $i_{eq}$, then $(x_{eq}, 0, i_{eq})$ is regarded as the equilibrium point where the nonlinear equation (\ref{dynamic}) is linearized.

We set $\bm z =(x_k-x_{eq}, \dot{x}_k, i_k-i_{eq})^\mathrm{T}$ as the state variable. Then, it follows from \cite{sun2016design} that  the linearized state space model of the magnetic levitation system can be expressed as:
\begin{eqnarray}
    \dot{\bm z} &=& A\bm z+B u \\
    y &=& C\bm z
\end{eqnarray}
where coefficient matrices $A$, $B$, and $C$ are given by

\begin{displaymath}
A =
\left[ \begin{array}{ccc}
0 & 1 & 0 \\
2450 & 0 & -1.1558 \\
0 & 2119.7 & -3.1438
\end{array} \right]
\end{displaymath}
\begin{figure}[t]
	\centerline{\includegraphics[width=70mm]{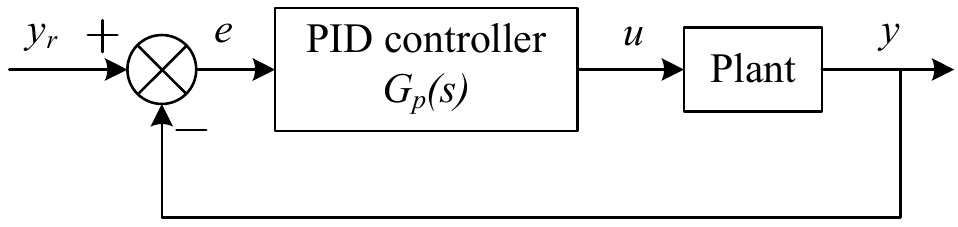}}
	\caption{Structure diagram of PID control.}
	\label{PID}
\end{figure}
\begin{equation}
\ B = \left[ \begin{array}{c}
0  \\
0  \\
2.6198
\end{array}
\right],
\ \
C = [\begin{array}{ccc}
1 & 0 & 0
\end{array}]
\label{linear model}
\end{equation}
and the output $y=x_k -x_{eq}$. The initial state is set to $\bm z_0 =(0.007, 0, 0)^\mathrm{T}$.

The PID controller is designed to solve the suspension regulation problem for the linearized model, see Fig. \ref{PID}. The transfer function of the PID controller is designed as
\begin{equation}
    G_p(s) = K_p + K_d s + K_i \frac{1}{s}
\end{equation}
where $K_p, K_d, K_i$ are the parameters of the PID controller.
\begin{figure}[t]
	\centerline{\includegraphics[width=70mm]{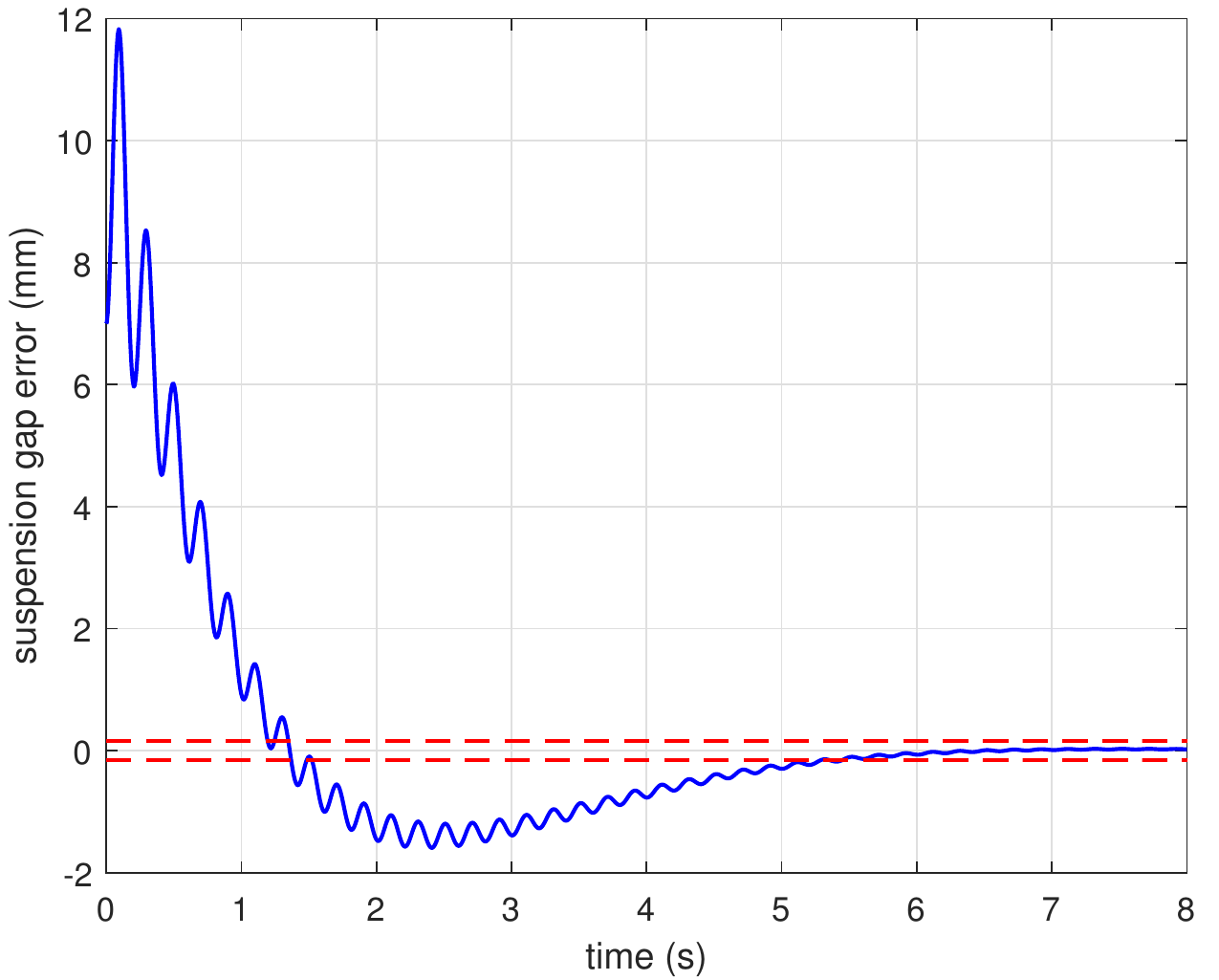}}
	\caption{Response of the suspension gap error using the PID controller.}
	\label{response}
\end{figure}

The differential time constant $\tau = K_d/K_p$ is set to $0.1$, and the proportional coefficient $K_p$ is obtained through the {\em root-locus method}. We set $K_p=3100$ in our simulation. The external disturbance $f(t)$ in (\ref{dynamic}) is generated by a Gaussian distribution $f(t) \sim \mathcal{N}(0, 2.5\times 10^5)$. To eliminate the static error induced by the external disturbance, we set $K_i=300$ empirically.

The response of the airgap error under the PID controller is shown in Fig. \ref{response}. We can observe that the air gap is finally regulated to the desired level with PID correction. Setting stability region is $2\%$, the suspension gap can reach a stable state in $5s$, and the maximum overshoot is about $25\%$. However, the error decreases with apparent oscillations and its setting time $5s$ is relatively too long for the suspension regulation problem.

\subsection{Linear Quadratic Gaussian Integral Control}
\begin{figure}[t]
\centerline{\includegraphics[,width=70mm]{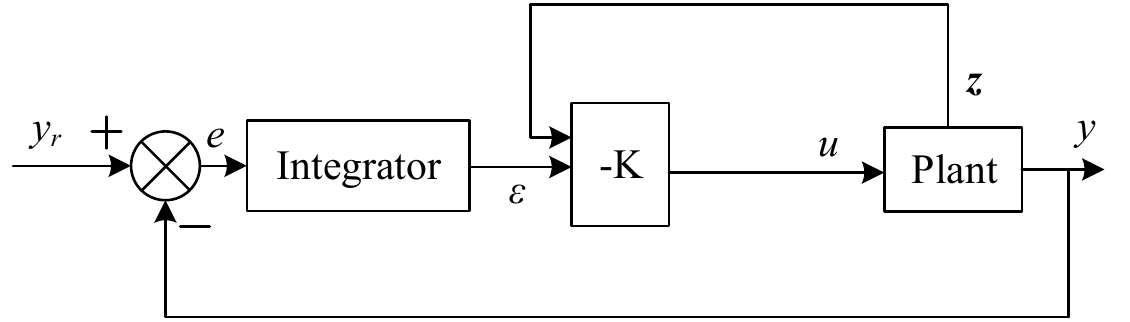}}
\caption{Structure diagram of LQI control.}
\label{lqi}
\end{figure}

The PID controller is widely applied for the magnetic levitation system of the maglev train, while it cannot address the inherent oscillation of the system, which significantly decelerates the convergence. The linear quadratic Gaussian integral (LQI) controller~\cite{young1972approach} is a full state feedback controller for the linear system, with the integral of the error as the additional input:
\begin{equation}
u = K[\bm{z} \  \epsilon]^T = K_{\bm{z}}\bm{z} + K_{\epsilon} \epsilon \\
\end{equation}
where $\epsilon$ is the output of the integrator
\begin{equation}
	\epsilon(t)= \int_{0}^{t} (y_r - y) dt.
\end{equation}

The gain matrix $K$ is the solution of an algebraic Riccati equation, which is derived from the minimization of the following cost function:
\begin{equation}
	J(u) = \int_{0}^{t}
	\left\{
		\left[\bm{z} \epsilon
		\right]Q
		\left[
		\begin{array}{c}
		\bm{z} \\
		\epsilon
		\end{array}
		\right]
			+Ru^2
	\right\}dt.
\end{equation}

The system applying the LQI controller does not have the issues encountered by the PID controller, i.e., the oscillation and the lack of robust. The LQI controller shown in Fig. \ref{lqi} is designed based on the linearized model (\ref{linear model}) of the magnetic levitation system. The response of the air gap error under LQI control is shown in Fig. \ref{rl_simu}. Notice that a model is indispensable in LQI control for the use of the algebraic Riccati equation. 

The regulation trajectory of LQI controller is displayed in Fig. \ref{rl_simu}. It is observed that its setting time ($0.1$s) is transparently much shorter than that of PID controller ($5$s) and there is almost no overshoot during the control phase.

\subsection{RL based controllers}
We discribe the experimental settings for the RL algorithm. Both DDPG and TD3 are implemented by Python 2.7 on linux system using Pytorch.

It should be noticed that the system model and the algorithm are implemented as discrete time version with sample time $dt = 1ms$. For example, the dynamical model for simulation is discretized using the forward Euler formula
\begin{equation}
    \bm z_{k+1} = \bm z_k + dt \cdot f(\bm z_k,  u_k).
\end{equation}

The episodes for every running are set to $M = 1000$. The time steps for each episode is set to $K = 500$. The structure of the neural networks are set precisely as described in Section \ref{sec_mdp}, in which each hidden layer contains $40$ neurons. After each time step, the networks are trained with a mini-batch of $100$ transitions, sampled uniformly from a replay buffer containing all the historical experiences whose size is $10^{6}$. The training of the networks begins after $10000$ time steps. The exploration noise of each action is set as Gaussain noise $\mathcal{N}(0,0.1)$. For the implementation of TD3, the target policy smoothing is implemented by adding $\epsilon \sim \mathcal{N}(0,0.2)$ to the actions generated by the target actor network, clipped to $(-0.5, 0.5)$. The actor network and the target critic networks update every $d = 2$ iterations.

\begin{figure}[t]
	\centerline{\includegraphics[width=70mm]{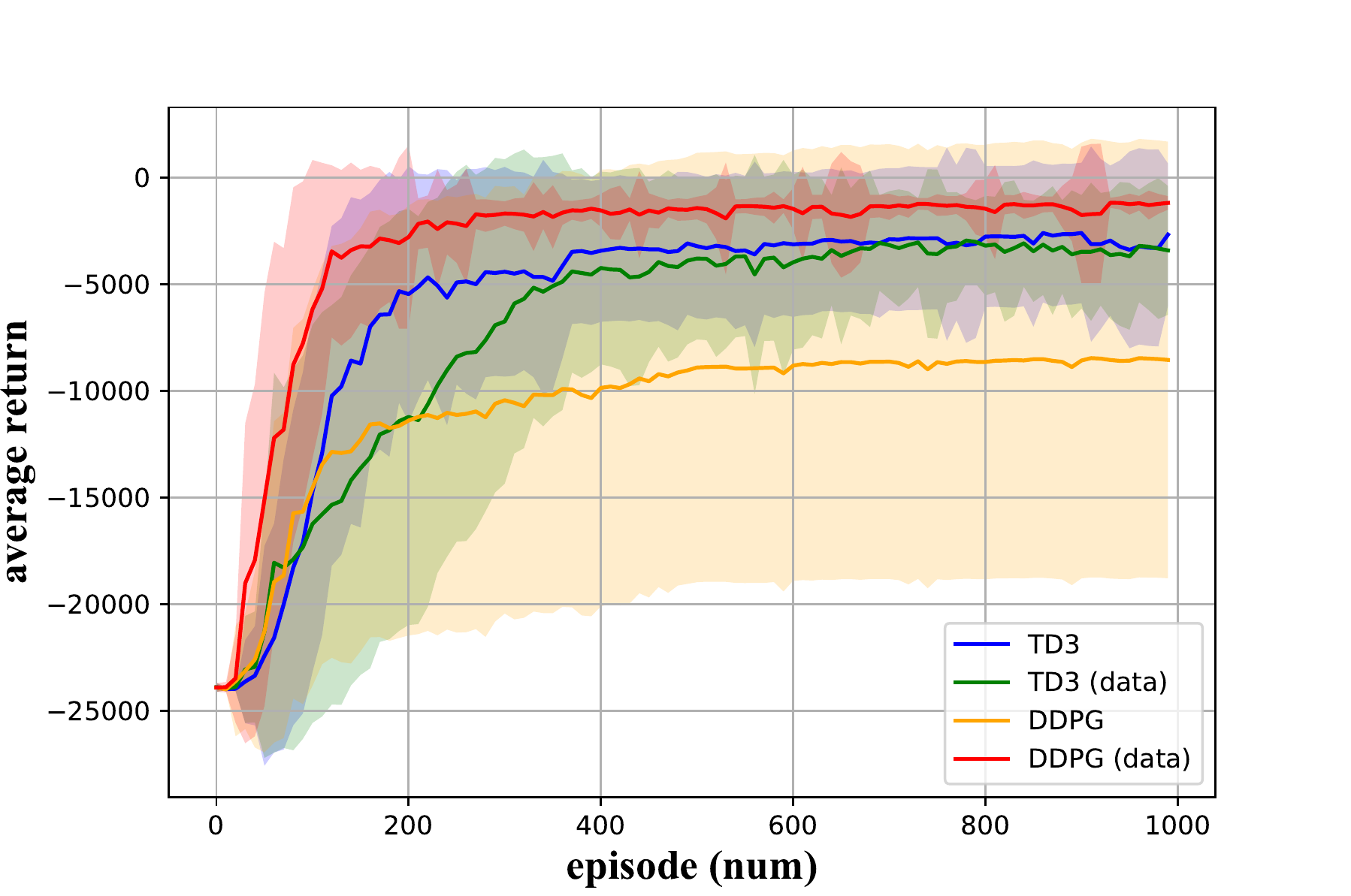}}
	\caption{Average return of proposed RL algorithms, with or without real samples filled into the replay cache at the beginning.}
	\label{return}
\end{figure}

The policy is evaluated every 5000 time steps and each evaluation reports the average reward over 10 episodes without exploration noise. Our results are reported over 50 random seeds of the environment and the network initialization. The average return of the proposed RL based controllers is shown in Fig. \ref{return}. We can easily observe that TD3 (blue line) converges faster than DDPG (orange line) and the stability is also improved significantly. 
\begin{figure}[t]
	\centerline{\includegraphics[width=70mm]{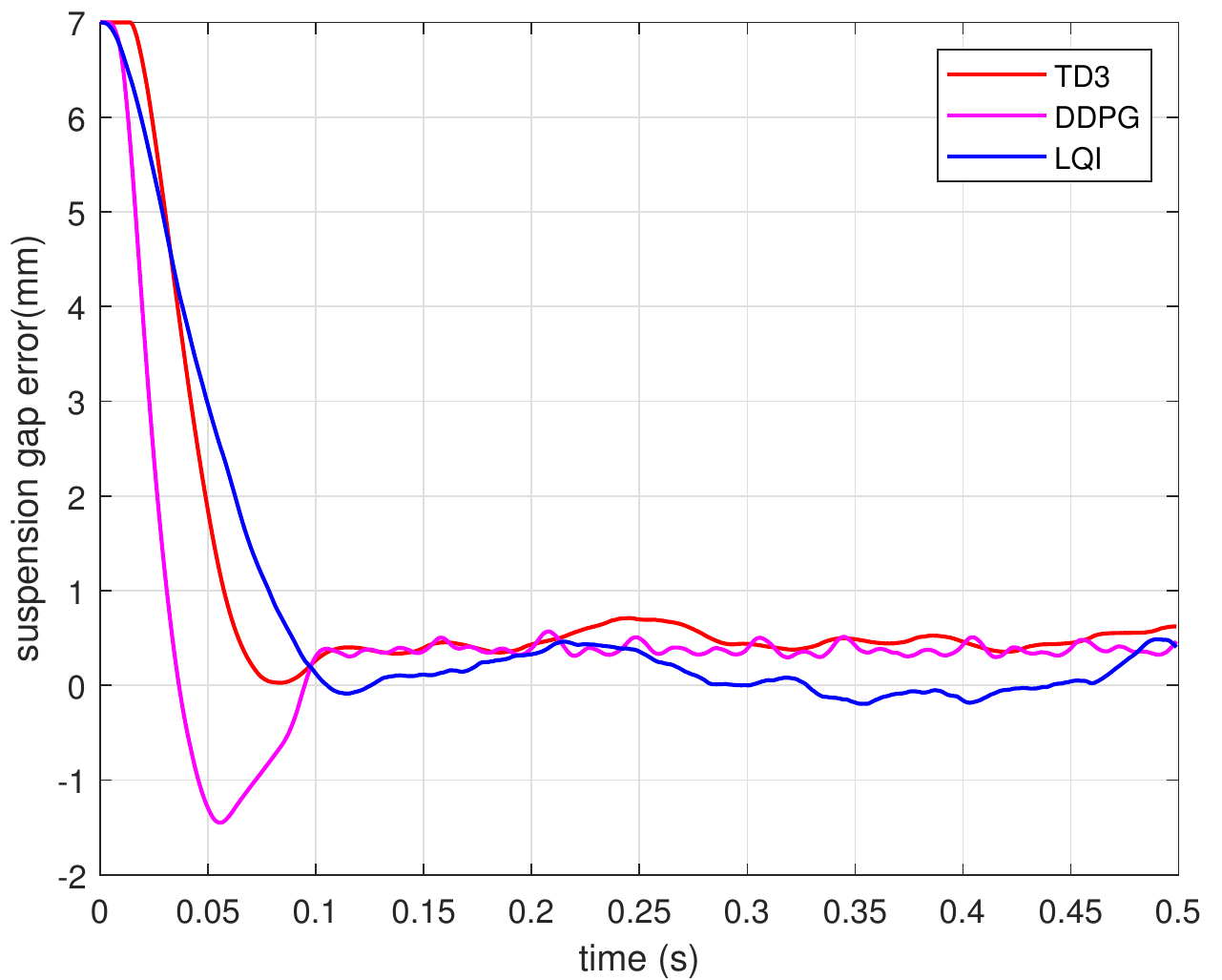}}
	\caption{Suspension gap error using TD3, DDPG and the LQI control.}
	\label{rl_simu}
\end{figure}
Then, we compare the performance of the RL based controller with those of PID and LQI on the suspension regulation problem. Fig. \ref{rl_simu} shows the regulation results of LQI and two RL based controllers from an initial air gap $x_0 = 15mm$ to a target air gap $x_{eq}=8mm$, where the trajectories of RL based controllers are generated using well-trained actor networks.

From the simulation, we can observe that PID behaves worst among all the controllers for its slow convergence and severe oscillations. In addition, the simulation shows that the performance of DDPG and TD3 is comparable to that of LQI, where an exact dynamical model is indispensable, and converge faster than the LQI.  While in the simulation of DDPG, we can find that there exists irregular oscillations in the whole sample horizon. An explanation is that the policy is approximated by the neural network. The overshoot of TD3 is trivial compared with DDPG, showing the improved performance of a double Q-learning framework. 
\begin{figure}[t]
	\centerline{\includegraphics[width=70mm]{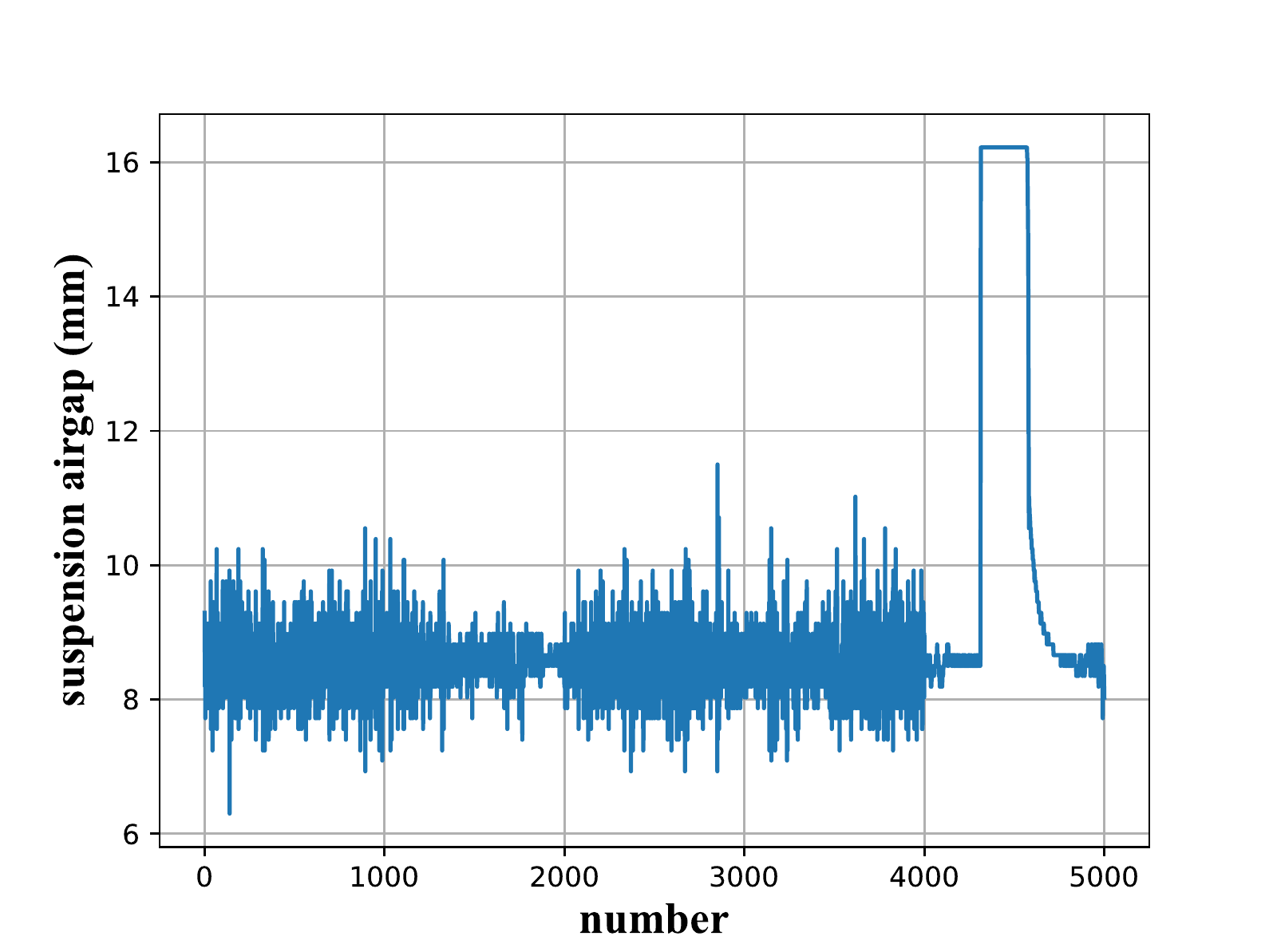}}
	\caption{Control trajectories of PID controller used in the real magnetic levitation system.}
	\label{data}
\end{figure}

\subsection{Learning from Real Dataset}
Moreover, we test the proposed RL framework using real samples obtained by the PID controller in the trial of mlsMTs in Changsha, China.

In Section \ref{sec_alg}, we propose an initialization method to reduce the sample complexity of RL algorithm, where the replay cache is initialized with samples from other controllers. The samples we used in our experiments are provided by CRRC using PID controller, and we carefully select valuable ones to balance the distribution of state-action pairs. Notice that the replay cache size is set to $10^6$, while we only store $10^4$ real samples, so that better experiences generated by the policy can affect more significantly as the actor network converges. Recall that a sample is a quadruple $(\bm s, \bm a, \bm s', c)$, where the cost is computed according to the environment. For simplicity, we only display a part of real control trajectories of the suspension airgap in Fig. \ref{data}. We believe that the RL algorithm learns faster at the beginning using better samples. 

It should be noticed that the sample time of the real dataset is $100$ms due to the communication limitation, while the minimum sample time to control the system effectively is conservatively estimated under $10$ms. This does not matter for our RL based controllers since RL algorithm values the distribution of the data more than its quality. Note that the mlsMT always run in the same rail so that the data is easily recyclable for the training of the neural networks. 

\begin{figure}[t]
\centerline{\includegraphics[width=70mm]{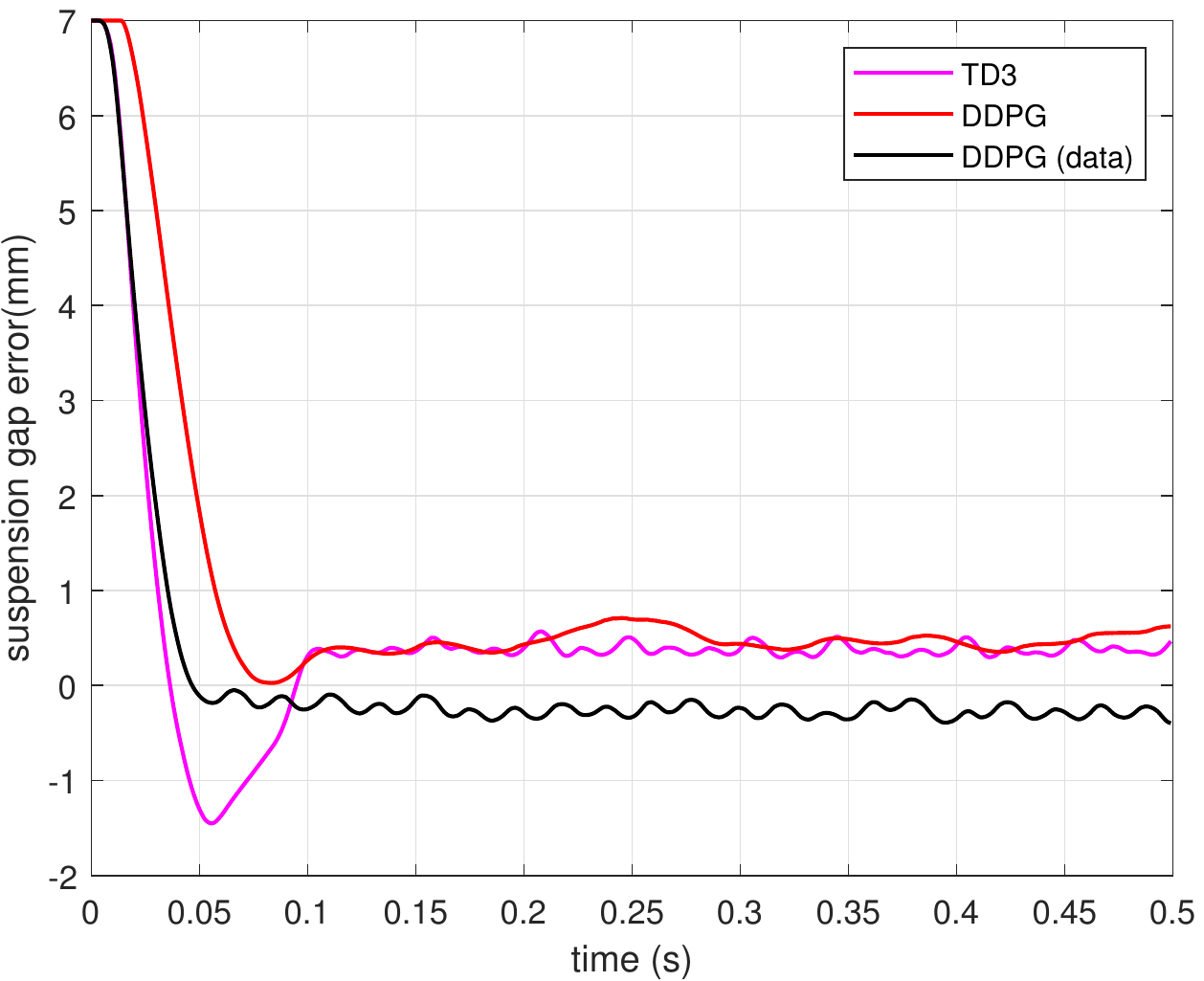}}
\caption{Suspension gap error using DDPG and TD3. DDPG (data) uses the real dataset for initialization.}
\label{rl_data}
\end{figure}

The average return in Fig. \ref{return} shows the tremendous improvement by learning from the real dataset. After using the real dataset, the learning efficiency of DDPG is greatly enhanced at the beginning and its variance is the smallest among the four methods. While we unexpectedly find that the TD3 with real samples performs worse. This phenomenon is termed as {\em negative transfer} in the transfer learning domain \cite{taylor2009transfer}. It is probably due to that the double Q-learning framework and the delayed update have addressed the underestimate errors of Q-value, thus TD3 prefers on-policy samples instead of off-policy ones like those from PID controller. We also observe that there exists a small steady-state bias (less than $0.5$mm, see both Fig.\ref{rl_simu} and Fig.\ref{rl_data}) on the suspension gap for our RL algorithms. The appearance of this bias occurs randomly as we repeatedly conduct the simulation with absolutely the same parameters, therefore it may be attributed to the randomness of RL algorithms, namely the random data and the neural networks. We compare the regulation results of DDPG with real data to that of pure DDPG and TD3 in Fig. \ref{rl_data}. The real samples help eliminating the overshoot of DDPG and the suspension gap error is reduced.

\section{Conclusion}\label{sec_con}
The paper proposed a sample-based RL framework for the suspension regulation of medium-low-speed maglev trains. We modeled the control problem as a continuous-state, continuous-action MDP with the well-designed state and one-step cost. The Q-value and policy are approximated by the neural networks. With the deterministic policy gradient, we proposed the DDPG algorithm for the suspension regulation that learns a feedback controller from past experiences. To improve the performance, we adopt a double Q-learning scheme to accelerate the training. We further discussed the algorithm implementation of the RL algorithm for the suspension regulation and used an initialization method to learn real dataset.

The  simulation results of the RL based controller were compared with model-based controllers, i.e., the PID controller and LQI controller. The performance of our controllers is better than PID and is even comparable with that of LQI based on an exact model. We also tested our algorithm using real dataset, and the performance of DDPG is significantly improved.

In the future, we will verify the proposed sample-based RL framework on the mlsMT in Changsha.

\bibliographystyle{IEEEtran}
\bibliography{mybibfile}
\end{document}